\documentclass[aps,prl,showpacs,twocolumn]{revtex4}
\usepackage{amssymb}
\usepackage{epsfig,graphicx}

\begin{document}

\title{Controllable Josephson-Like Tunneling in Two-Component Bose-Einstein
Condensates Coupled with Microwave via Feshbach Resonance and Trapping
Potential}
\author{Bo Xiong$^{1,2}$, Weiping Zhang$^{2}$, W. M. Liu$^{1}$}
\affiliation{$^1$Beijing National Laboratory for Condensed Matter Physics, Institute of
Physics, Chinese Academy of Sciences, Beijing 100080, China}
\affiliation{$^2$Key Laboratory of Optical and Magnetic Resonance Spectroscopy (Ministry
of Education), Department of Physics, East China Normal University, Shanghai
200062, China}
\date{\today}

\begin{abstract}
We put forward a scheme for controlling Josephson-like tunneling in
two-component Bose-Einstein condensates coupled with microwave field via
Feshbach resonance and tuning aspect ratio of trapping potential. We prove
how to realize a perfect periodic oscillation from a fast damped and
irregular oscillation on relative number of atoms in future experiment. In
particular, intensity of Josephson-like tunneling can be successfully
controlled through controlling speed of recovering the initial value of
intra-atomic interaction and aspect ratio of trapping potential.
Interestingly, we find that relative number of atoms represents two
different types of oscillation in respond to periodic modulation of
attractive intra-atomic interaction.
\end{abstract}

\maketitle

The existence of a Josephson current is a direct manifestation of
macroscopic quantum phase coherence \cite{Barone1982} and have numerous
important applications in condensed matter physics, quantum optics and cold
atom physics, for example, precision measurement, quantum computation. The
physical origin of the Josephson current is the temporal interference of the
two systems which must both have a well defined quantum phase and a
different average energy per particle, respectively. Recently, the
experimental realization of multi-component Bose-Einstein condensates (BECs)
of weakly interacting alkali atoms has provided a route to study\ Josephson
effect \cite%
{Hall1998-1,MatthewsPRL1998,LeggetRMP,AndersonScience,WilliamsPRA,ParkPRL,CataliottiScience,KohlerPRL,ShinPRL}
in a controlled and tunable way by means of Feshbach resonance \cite%
{SaitoPRL,KevrekidisPRL,PelinovskyPRL,Perez-GarciaPRL,KonotopPRL,MatuszewskiPRL}
so far unattainable in charged systems. The physical origin of the
low-energy Feshbach resonances is the\textbf{\ }low-energy binary collisions
described by the difference of the initial and intermediate state energies
which can be effectively altered through variations of the strength of an
external magnetic field.

In this Letter, we put forward a scheme for controlling Josephson-like
tunneling in two-component BECs coupled with microwave field via Feshbach
resonance and tuning aspect ratio of trapping potential. we find that,
through time-dependent tuning attractive intra-atomic interaction, a
perfectly periodic oscillation can be obtained successfully. Especially, by
controlling speed of recovering the initial value of attractive intra-atomic
interaction and aspect ratio of trapping potential, we can successfully
control intensity of Josephson-like tunneling. Furthermore, relative number
of atoms represents two different types of oscillation in respond to
periodic modulation of attractive intra-atomic interaction.

We consider a system of longitudinal elongated two-component BECs coupled by
the microwave field with the effective Rabi frequency $\Omega $ and finite
detune $\delta $, where the $\Omega $ and $\delta $ are independent upon
time and space coordinate as experimental case \cite{MatthewsPRL1998}. At
zero temperature, this system can be described by macroscopic wave function $%
\psi _{j}$\ (normalized to unity, i.e., $\int (|\psi _{1}|^{2}+|\psi
_{2}|^{2})dz=1$) which obey the dimensionless one-dimensional
Gross-Pitaevskii equations (GPEs) as \cite{WilliamsPRA,ParkPRL}%
\begin{eqnarray}
i\frac{\partial }{\partial t}\psi _{j}(z,t) &=&[-\frac{\partial ^{2}}{%
\partial z^{2}}+V(z)]\psi _{j}(z,t)+  \nonumber \\
&&[G_{jj}(t)|\psi _{j}|^{2}+G_{jk}(t)|\psi _{k}|^{2}]\psi _{j}(z,t)+
\nonumber \\
&&(-1)^{j+1}\frac{\delta }{\omega _{\bot }}\psi _{j}(z,t)+\frac{\Omega }{%
\omega _{\bot }}\psi _{k}(z,t),  \label{GP2}
\end{eqnarray}%
where $j,k=1,2$, $j\neq k$, and $\omega _{\bot }$ is radial harmonic
trapping frequency, time and coordinate unit are $2/\omega _{\bot }$ and $%
a_{\bot }=\sqrt{\hbar /m\omega _{\bot }}$, respectively, and $V(z)$ is the
longitudinal confining potential, $G_{jj}=(4N\hbar a_{jj})/(ma_{\bot
}^{3}\omega _{\bot })$ and $G_{jk}=(4N\hbar a_{jk})/(ma_{\bot }^{3}\omega
_{\bot })$ are dimensionless intra- and inter-species effective atomic
interaction potentials, respectively, where $a_{jj}$ and $a_{jk}$ are
corresponding s-wave scattering lengths, respectively, $N$ is the total
atomic number, and $m$ is the atomic mass assumed the same for both two
species.

In the following discussion, we consider two-hyperfine spin states of $%
^{7}Li $ bose atom which the magnitude and sign of the atomic interaction
can be tuned to any value via a magnetic-field dependent Feshbach resonance.
And we choose dimensionless $V(z)=\gamma ^{2}z^{2}$, where $\gamma =\omega
_{z}/\omega _{\bot }$ is aspect ratio of trapping potential, $\omega _{z}$
is longitudinal harmonic trapping frequency and $\omega _{\bot }=2\pi \times
625$ Hz (so, the time and coordinate units are $0.51$ ms and $a_{\bot }=1.51$
$\mu $m respectively). Moreover, considering the condensates will be
unstable and collapse when the number of particles is large enough \cite%
{Muryshev}, so we choose $N=6000$, which provides a safe range of parameters
for avoiding instability and collapse occurring. And the effective Rabi
frequency and finite detune are chosen as $\Omega =1.5\omega _{\bot }$ and $%
\delta =0.5\omega _{\bot }$ considering experimental feasibility.
\begin{figure}[tbp]
\epsfxsize=9.5cm \centerline{\epsffile{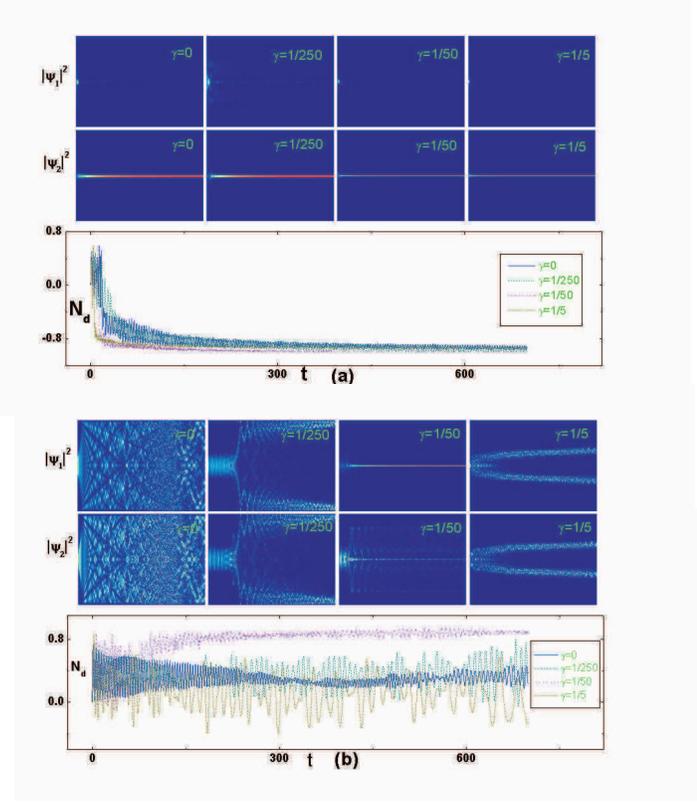}}
\caption{(Color online) Temporal evolution of atomic density and relative
number of atoms $N_{d}$ for initial Gaussian wave function in response to
various aspect ratio of trapping potential $\protect\gamma $. (a)
Corresponding to attractive atomic interactions for the parameters as $%
G_{11}=G_{22}=4G_{12}=-10$. (b) Corresponding to repulsive atomic
interactions for the parameters as $G_{11}=G_{22}=4G_{12}=10$. }
\end{figure}

First of all, we consider the simplest case, i.e., intra- and inter-species
effective atomic interaction potentials are\ time-independent attractive or
repulsive case, which can be easily realized using Feshbach resonance
experimentally. Here, we have chosen $a_{11}=a_{22}$ is just considering the
experimental feasibility and assumed that tuning atomic interaction wouldn't
destroy the condensates in this system. We denote relative number of atoms
between two species as $N_{d}$, i.e., $N_{d}=N_{1}-N_{2}$, where $N_{1}=\int
|\psi _{1}|^{2}dz$ and $N_{2}=\int |\psi _{2}|^{2}dz$, respectively. We
study Josephson-like tunneling in this case and the results are shown in
Fig. 1, which is based on numerical simulation of Eqs. (\ref{GP2}) for the
initial Gaussian wave function under the various aspect ratio of trapping
potential. As shown in Fig. 1a corresponding to attractive atomic
interactions, during the evolution process, exchange of atoms is always
damped very fast and almost suppressed completely irrespective of aspect
ratio of trapping potential. But, on the contrary, in repulsive atomic
interactions corresponding to Fig. 1b, exchange of atoms vary significantly
in response to various aspect ratio of trapping potential. Detailed analyses
of above phenomena will be given in the following.
\begin{figure}[tbp]
\epsfxsize=9.5cm \centerline{\epsffile{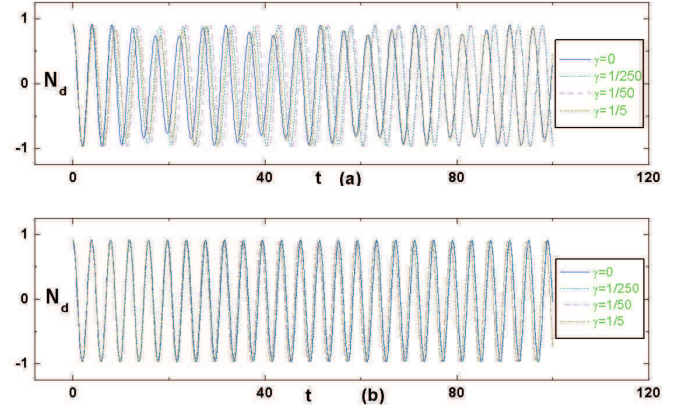}}
\caption{(Color online) Temporal evolution of $N_{d}$ for initial Gaussian
wave function in response to various aspect ratio of trapping potential $%
\protect\gamma $, which is based on numerical simulation of Eqs. (\protect
\ref{GP3}) for the case $\Sigma (\protect\eta )=0$. (a) Corresponding to
attractive atomic interactions for the parameters as $%
G_{11}=G_{22}=4G_{12}=-10$. (b) Corresponding to repulsive atomic
interactions for the parameters as $G_{11}=G_{22}=4G_{12}=10$. }
\end{figure}

From Eqs. (\ref{GP2}) and taking boundary condition into account, it is not
difficult to derive following equations,

\begin{equation}
\frac{dN_{d}}{dt}=\frac{4\Omega }{\omega _{\bot }}\int (\sqrt{\rho _{1}\rho
_{2}}\sin \phi _{d})dz,  \label{AD}
\end{equation}%
where wave functions have been chosen as $\psi _{1}=\sqrt{\rho _{1}}\exp
(i\phi _{1})$, $\psi _{2}=\sqrt{\rho _{2}}\exp (i\phi _{2})$, respectively,
and $\phi _{d}=\phi _{1}-\phi _{2}$. From Fig. 1a, we can see that maximum $%
\rho _{1}$ will be much smaller than maximum $\rho _{2}$ very quickly for
all aspect ratio of trapping potential, so maximum $\sqrt{\rho _{1}\rho _{2}}
$ will decrease rapidly. Moreover, if considering well localized density
profiles of two species as shown in Fig. 1a, from Eqs. (\ref{AD}), we can
immediately conclude that amplitude of oscillation of $N_{d}$ will be damped
very fast and suppressed completely in this case. But from Fig. 1b
corresponding to repulsive atomic interactions, we can see that the atomic
densities of two species vary significantly in response to various aspect
ratio of trapping potential and represent nonlocalized profiles, so time
evolution of amplitudes of $N_{d}$ represents very different behavior from
attractive case.

As pointed out above, $N_{d}$ will represent a very irregular and fast
damped oscillation in time-independent attractive atomic interactions and
represent a more irregular oscillation in time-independent repulsive case.
But, from the experimental point of view, valuable and applicable systems
usually have regular and stable characteristics rather than irregular and
unstable one. Fortunately, we find out that, through time-dependent tuning
atomic interaction via Feshbach resonance, regular and stable oscillation of
$N_{d}$ can be successfully realized in the attractive case, but not be
realized in the repulsive case. Now, let us to understand physics from both
theoretical analysis and numerical simulation aspects.
\begin{figure}[tbp]
\epsfxsize=9.5cm \centerline{\epsffile{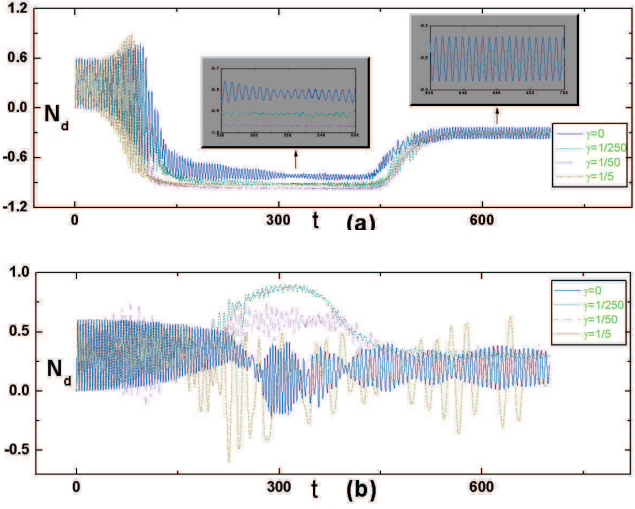}}
\caption{(Color online) Temporal evolution of relative number of atoms $N_{d}
$ for initial Gaussian wave function in response to various aspect ratio of
trapping potential $\protect\gamma $. (a) Corresponding to attractive atomic
interactions for the parameters as $G_{12}=-2.5$, $\protect\tau =105$, $%
t_{1}=0.5t_{2}=280$. (b) Corresponding to repulsive atomic interactions for
the parameters as $G_{12}=2.5$, $\protect\tau =105$, $t_{1}=0.5t_{2}=280$. }
\end{figure}

In order to study how to control Josephson-like tunneling in this system, we
diagonalize linear parts of Eqs. (\ref{GP2}) by means of applying an unitary
transformation $U$ on both sides of Eqs. (\ref{GP2}), then it is transformed
into following form,

\begin{eqnarray}
i\frac{\partial }{\partial t}\varphi _{\pm }(z,t) &=&[-\frac{\partial ^{2}}{%
\partial z^{2}}+V(z)]\varphi _{\pm }(z,t)+\zeta _{1}^{\pm }|\varphi _{\pm
}|^{2}\varphi _{\pm }(z,t)  \nonumber \\
&&+\zeta _{2}^{\pm }|\varphi _{\mp }|^{2}\varphi _{\pm }(z,t)+\Sigma (\eta ),
\label{GP3}
\end{eqnarray}%
where%
\[
U\left(
\begin{array}{c}
\psi _{1} \\
\psi _{2}%
\end{array}%
\right) \!\!=\!\!\left(
\begin{array}{c}
\varphi _{+}e^{i\eta t} \\
\varphi _{-}e^{-i\eta t}%
\end{array}%
\right) \!\equiv \!\left(
\begin{array}{c}
\Psi _{+} \\
\Psi _{-}%
\end{array}%
\right) \!,\!U\!=\!\left(
\begin{array}{cc}
\sin \!\theta \! & \!-\cos \!\theta \\
\cos \theta \! & \!\sin \theta%
\end{array}%
\right) \!,\!
\]%
\begin{eqnarray*}
\Sigma (\eta ) &=&\zeta _{3}^{\pm }\varphi _{\pm }^{2}\varphi _{\mp }^{\star
}e^{\pm 2i\eta t}+\zeta _{4}^{\pm }|\varphi _{\pm }|^{2}\varphi _{\mp
}e^{\mp 2i\eta t} \\
&&+\zeta _{5}^{\pm }\varphi _{\mp }^{2}\varphi _{\pm }^{\star }e^{\mp 4i\eta
t}+\zeta _{6}^{\pm }|\varphi _{\mp }|^{2}\varphi _{\mp }e^{\mp 2i\eta t},
\end{eqnarray*}%
here $\zeta _{1}^{\pm }=[(3G_{11}+2G_{12}+3G_{22})\mp 4(G_{11}-G_{22})\cos
2\theta +(G_{11}-2G_{12}+G_{22})\cos 4\theta ]/8$, $\zeta _{2}^{\pm
}=[(G_{11}+2G_{12}+G_{22})-(G_{11}-2G_{12}+G_{22})\cos 4\theta ]/4$, $\zeta
_{3}^{\pm }=[2(G_{11}-G_{22})\sin 2\theta \mp (G_{11}-2G_{12}+G_{22})\sin
4\theta ]/8$, $\zeta _{4}^{\pm }=[2(G_{11}-G_{22})\sin 2\theta \mp
(G_{11}-2G_{12}+G_{22})\sin 4\theta ]/4$, $\zeta _{5}^{\pm
}=[(G_{11}-2G_{12}+G_{22})\sin ^{2}2\theta ]/4$, $\zeta _{6}^{\pm
}=[2(G_{11}-G_{22})\sin 2\theta \pm (G_{11}-2G_{12}+G_{22})\sin 4\theta ]/8$%
, and $\theta =(1/2)\tan ^{-1}\left( \Omega /\delta \right) $, $\eta
=[\left( \delta /\omega _{\bot }\right) ^{2}+\left( \Omega /\omega _{\bot
}\right) ^{2}]^{1/2}$. Now, we can see that, comparing with the original
Eqs. (\ref{GP2}), we introduce a new term denoted by $\Sigma (\eta )$ which
is exponential time dependence, and it is not included in regular coupling
nonlinear Schr\"{o}dinger equations. It must be pointed out that $\Sigma
(\eta )$ depends not only on the effective atomic interaction potentials $%
G_{jk}$ $(j$,$k=1$,$2)$, but on the effective Rabi frequency $\Omega $ and
finite detune $\delta $, i.e., $\theta $ and $\eta $.
\begin{figure}[tbp]
\epsfxsize=9.5cm \centerline{\epsffile{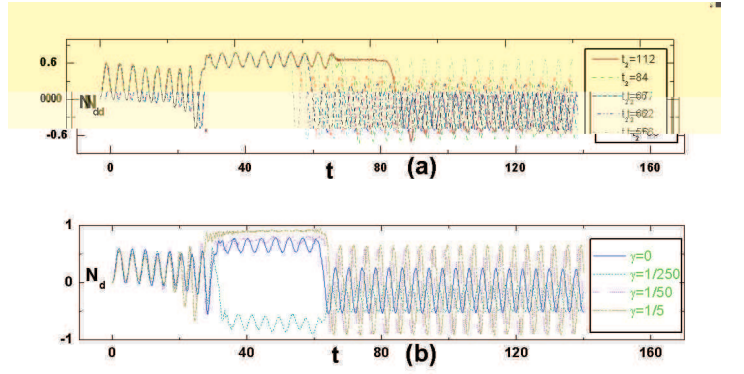}}
\caption{(Color online) Temporal evolution of $N_{d}$ for initial Gaussian
wave function in response to (a) various speed of recovering the initial
value of intra-atomic interaction for the parameters as $\protect\gamma =0$,
$G_{12}=-2.5$, $\protect\tau =21$, $t_{1}=56$, and (b) various aspect ratio
of trapping potential for the parameters as $G_{12}=-2.5$, $\protect\tau =21$%
, $t_{1}=0.83t_{2}=56$.}
\end{figure}

From the expression of $\Sigma (\eta )$, we can find that, if $\varphi _{\pm
}(z,t)$ vary slowly over time period $\pi /\eta $, then the effects of $%
\Sigma (\eta )$ on Josephson-like tunneling can be averaged out over time
period $\pi /\eta $ during the evolution process. But if $\varphi _{\pm
}(z,t)$ vary significantly over time period $\pi /\eta $, then $\Sigma (\eta
)$ will strongly affect Josephson-like tunneling. In the following, we will
investigate effects of $\Sigma (\eta )$ on Josephson-like tunneling in
detail.

Firstly, we consider the special case that $\Sigma (\eta )=0$. From Eqs. (%
\ref{GP3}) and transformation relation between original and transformed wave
function, we can obtain that $N_{d}$ will represent a large-amplitude
periodic oscillation in both attractive and repulsive atomic interactions.
Above conclusions are well confirmed by Fig. 2, which is based on numerical
simulation of Eqs. (\ref{GP3}) and then applying inverse unitary
transformation $U^{-1}$ on $\Psi _{\pm }$ in order to obtain original wave
function. So, it is reasonable to infer that the reasons for fast damped and
irregular oscillation of $N_{d}$ are mainly due to the effects of $\Sigma
(\eta )$ on Josephson-like tunneling.

But, key question is how to efficiently suppress effects of $\Sigma (\eta )$
on Josephson-like tunneling in order to obtain perfectly periodic
oscillation of $N_{d}$. In the following, we prove that, only in the
attractive case, through time-dependent tuning intra-atomic interaction via
Feshbach resonance, we can successfully control the effects of $\Sigma (\eta
)$ and obtain a perfectly periodic oscillation on relative number of atoms.
Considering the experimental feasibility, in the following case, we choose
time-dependent intra-atomic interaction as \cite%
{InouyeNature,RobertsPRL,StreckerNature}
\begin{equation}
G_{jj}(t)=\left\{
\begin{array}{c}
G_{12}\exp (t/\tau ), \\
G_{12}\exp [(t_{2}-t)/\tau ], \\
G_{12},%
\end{array}%
\right.
\begin{array}{c}
0\leqslant t\leqslant t_{1}; \\
t_{1}<t\leqslant t_{2}; \\
t>t_{2}.%
\end{array}
\label{FR}
\end{equation}%
where $j=$($1$, $2$), $G_{11}(t)=G_{22}(t)$ and $G_{12}$ is
time-independent parameter determining inter-atomic interaction,
$\tau $ determines speed of tuning intra-atomic interaction
depending on the values of $t_{1}$ and $t_{2} $ at fixed minimum
$G_{jj}(t)$. The results are shown in Fig. 3. It is very interesting
that $N_{d}$ finally represents a perfectly periodic oscillation in
the attractive case, but not in the repulsive case. Above phenomena
are mainly due to successfully suppress the effects of $\Sigma (\eta
)$ on Josephson-like tunneling in the attractive case, but not in
the repulsive case, which underlying physics is just because
$\varphi _{\pm }(z,t)$ vary significantly over time period $\pi
/\eta $ in the repulsive case. So, in the following, we will
concentrate on attractive case. It is important to point out that,
in the attractive case, the amplitude of $N_{d}$ is strongly
dependent upon aspect ratio of trapping potential as shown in Fig.
3a. Based upon the above analysis, we put forward a scheme for
controlling the amplitude of oscillation through controlling speed
of recovering the initial value of intra-atomic interaction (i.e.,
changing value of $t_{2}$ but fixing other parameters in Eqs.
(\ref{FR})) and tuning aspect ratio of trapping potential.

The results are shown in Fig. 4. It is very interesting to note that
oscillatory amplitude of $N_{d}$ can be controlled successfully not only via
Feshbach resonance but tuning aspect ratio of trapping potential. Above
phenomena can be well understood as follows: If $\varphi _{\pm }(z,t)$ vary
slowly over time period $\pi /\eta $, then the effects of $\Sigma (\eta )$
on Josephson-like tunneling can be effectively suppressed, and $N_{d}$ will
represent a perfectly periodic oscillation which amplitude is determined by
atomic density of two species.
\begin{figure}[tbp]
\epsfxsize=9.5cm \centerline{\epsffile{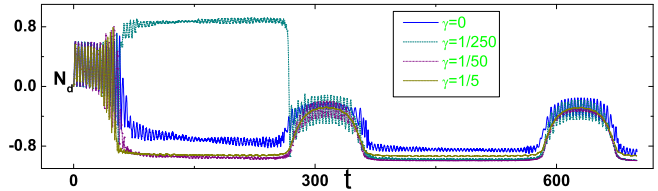}}
\caption{(Color online) Temporal evolution of relative number of atoms $N_{d}
$ for initial Gaussian wave function in response to periodic modulation of
attractive intra-atomic interaction and various aspect ratio of trapping
potential for the parameter as $G_{12}=-2.5$.}
\end{figure}

We also study Josephson-like tunneling in response to periodic modulation of
attractive interaction \cite{KevrekidisPRL} and various aspect ratio of
trapping potential. In this case, attractive atomic interactions have been
chosen as $G_{11}(t)=G_{22}(t)=G_{12}[1+8\sin ^{2}(0.01t)]$, where $%
G_{12}=-2.5$. As shown in Fig. 5, it is very interesting to note that $N_{d}$
represents two different types of oscillation, which one represents small
frequency behavior originated from periodic modulation of attractive
interaction, but the other represents large frequency behavior originated
from coupling microwave. Above effects provide a promising protocol to
realize a new type of switch controlled by periodically managed Feshbach
resonance in future experiments.

In conclusion, we prove how to successfully control Josephson-like tunneling
in attractive two-component BECs coupled by microwave field via Feshbach
resonance in respond to various aspect ratio of trapping potential. Recent
developments of controlling the scattering length and realization of
multi-component BECs in the experiments allow for the experimental
investigation of our prediction in future.

This work was supported by the NSF of China under 10474055, 90406017,
10588402, 60525417; the NKBRSF of China under 2005CB724508; the STCM of
Shanghai under 05PJ14038, 06JC14026, 04DZ14009.


\begin{thebibliography}{99}
\bibitem{Barone1982} A. Barone, G. Paterno, Physics and Applications of the
Josephson Effect (Wiley, New York, 1982).

\bibitem{Hall1998-1} D. S. Hall et al., Phys. Rev. Lett. \textbf{81}, 1539
(1998); ibid \textbf{81}, 1543 (1998).

\bibitem{MatthewsPRL1998} M. R. Matthews et al., Phys. Rev. Lett. \textbf{81}%
, 243 (1998); ibid \textbf{83}, 3358 (1999).

\bibitem{LeggetRMP} A. J. Legget, Rev. Mod. Phys. \textbf{73}, 307 (2001).

\bibitem{AndersonScience} B. P. Anderson et al., Science \textbf{282}, 1686
(1998).

\bibitem{WilliamsPRA} J. Williams et al., Phys. Rev. A \textbf{59}, R31
(1999).

\bibitem{ParkPRL} Q. H. Park et al., Phys. Rev. Lett. \textbf{85}, 4195
(2000).

\bibitem{CataliottiScience} F. S. Cataliotti et al., Science \textbf{293},
843 (2001).

\bibitem{KohlerPRL} S. Kohler et al., Phys. Rev. Lett. \textbf{89}, 060403
(2002).

\bibitem{ShinPRL} Y. Shin et al., Phys. Rev. Lett. \textbf{95}, 170402
(2005).

\bibitem{SaitoPRL} H. Saito et al., Phys. Rev. Lett. \textbf{90}, 040403
(2003).

\bibitem{KevrekidisPRL} P. G. Kevrekidis et al., Phys. Rev. Lett. \textbf{90}%
, 230401 (2003).

\bibitem{PelinovskyPRL} D. E. Pelinovsky et al., Phys. Rev. Lett. \textbf{91}%
, 240201 (2003).

\bibitem{Perez-GarciaPRL} V\'{\i}ctor M. P\'erez-Garc\'{\i}a et al., Phys. Rev. Lett.
\textbf{92}, 220403 (2004).

\bibitem{KonotopPRL} V. V. Konotop et al., Phys. Rev. Lett. \textbf{94},
240405 (2004).

\bibitem{MatuszewskiPRL} M. Matuszewski et al., Phys. Rev. Lett. \textbf{95}%
, 050403 (2005).

\bibitem{Muryshev} A. E. Muryshev et al., Phys. Rev. Lett. \textbf{60},
R2665 (1999).

\bibitem{InouyeNature} S. Inouye et al., Nature \textbf{392}, 151 (1998).

\bibitem{RobertsPRL} J. L. Roberts et al., Phys. Rev. Lett. \textbf{85}, 728
(2000).

\bibitem{StreckerNature} K. E. Strecker, et al., Nature \textbf{417}, 150
(2002).
\end{thebibliography}
\end{document}